\documentstyle[aas2pp4,amssym,epsfig]{article}

\righthead{12\,$\mu$m Seyferts}
\lefthead{Hunt, L.K. et al.}

\begin{document}

\title{Morphology of the 12\,$\mu$m Seyfert Galaxies:
I. Hubble Types, Axial Ratios, Bars, and Rings} 

\author{ L. K. Hunt}
\affil{ C. A. I. S. M. I. - C. N. R. \\
Largo E. Fermi 5, I-50125 Firenze, Italy\\
Electronic mail: hunt@arcetri.astro.it}

\and

\author { M. A. Malkan}
\affil{ University of California \\
Department of Astronomy, 405 Hilgard Ave., \\ 
Los Angeles, CA, 90095-1562 U.S.A. \\
Electronic mail: malkan@bonnie.astro.ucla.edu}

\begin{abstract}
We have compared the morphological characteristics of the 891 galaxies in
the Extended 12\,$\mu$m Sample (E12GS), and assessed the effect of the
12\,$\mu$m selection criterion on galaxy properties.
The normal spirals in the E12GS
have the same axial ratios, morphological types, and bar and ring
fractions as other normal spirals. The HII/starburst galaxies have a higher
incidence of bars, and more than twice the normal rate of ``peculiar"
morphologies, both of which are attributable to relatively recent
disturbances.  

The 12\,$\mu$m Seyferts show a small (10\%) deficiency of edge-on disks.
This is caused by extinction, but is a much less severe effect than
in optically-selected samples.  There is a similar modest deficit of
highly inclined HII/starburst galaxies in the 12\,$\mu$m sample.

The galaxies with active nuclei (AGNs: Seyferts and LINERs) have the
same incidence of bars as normal spirals,  
but show rings significantly more
often than normal galaxies or starbursts. The LINERs have elevated rates of
{\it inner} rings, while the Seyferts have {\it outer} ring fractions
several times those in normal galaxies.  
The different formation times of bars and rings suggest an interpretation
of these differences.
Bars form relatively quickly, and indicate that material is
recently being transported (by redistribution of angular momentum)
to the center of the galaxy, where it
is likely to trigger a short (e.g., $\lesssim\,10^8$ yrs) burst of star formation.  
Outer rings may result from similar disturbances, but require much
more time to form.
They would then be associated
with more intense nuclear activity if it takes $10^9$ years or more for the mass
transfer to reach the center and raise the black hole accretion rate, by
which time the bar has dissolved or begun to do so.
Inner rings form before outer ones, with a formation time more comparable to bars.
Thus it may be that after an interaction or instability triggers an
infall of gas, the galaxy in the earliest stage is likely to show enhanced
star formation in its center, while later it is more likely to show
LINER activity, and still later likely to be a Seyfert. 

The trends we find with morphology and nuclear activity are not biased 
either by the distances of the galaxies, or by the slightly elevated 
recent star formation rates 
shown by the 12\,$\mu$m galaxies in general. 
\end{abstract}


\section{Introduction}

While for many years it was thought that initial conditions 
uniquely determined galaxy morphology (Eggen, Lynden-Bell, \& Sandage \cite{els}), 
it is now becoming apparent that morphology can be modified by
physical processes (e.g., Pfenniger, Combes, \& Martinet \cite{pcm}).
It follows that galaxy morphology can be used to study these
processes, if a relationship can be established between a
morphological feature and the physical mechanism responsible. 

The most striking features of disk galaxy morphology are nonaxisymmetric 
structures such as bars and spiral patterns. 
Such structures can be caused by instabilities in galactic disks, and 
the interaction between bars and disks or bulges can give rise to 
angular momentum transfer and resonance phenomena. 
Gas, because of its dissipative properties, is expected to have a substantial
influence on the development of spiral structure and bars.
Numerical simulations suggest that bars can induce substantial inflows of gas
(Schwarz \cite{schwarz}; Friedli \& Benz \cite{friedli:benz93}), 
and the consequent evolution of the bar may create thickened structures that
resemble bulges
(Norman, Sellwood, \& Hasan \cite{norman}; Friedli \& Benz \cite{friedli:benz}).
Environmental effects, such as
tidal interactions and mergers, can also induce bars, as well as bridges, 
tails, multiple nuclei, and highly non-axisymmetric ``disturbed'' structure.
Interactions may also alter the Hubble type in spirals, causing them to evolve
from late-type unbarred systems 
toward barred earlier types (Elmegreen, Elmegreen, \& Bellin \cite{eeb}).

Axisymmetric structures in galaxies, bulges and disks, 
may also be modified over time by these processes and others, 
and the Hubble sequence itself may turn out to
be an evolutionary one (Pfenniger, Combes, \& Martinet \cite{pcm}; 
Martinet \cite{martinet}).
If this were true, there could be a link between normal spiral evolution
and the triggering of starburst and nuclear (Seyfert) activity.
These kinds of activity would be expected if such evolution,
instead of proceeding at quiescent quasi-static rates, were to occur
violently on relatively short timescales, and involve only modest 
fractions of the central mass.

In this paper,
we investigate the morphology of several classes of disk galaxies
in an attempt to better understand the physical processes behind the creation
and maintenance of an active galactic nucleus (AGN).
Much effort has been devoted to identifying the connection, if any, between
host galaxy properties and the AGN
(Simkin, Su, \& Schwarz \cite{sss}; Su \& Simkin \cite{ss};
Yee \cite{yee}; MacKenty \cite{mackenty}; Zitelli et al. \cite{zitelli};
Danese et al. \cite{danese}; Granato et al. \cite{granato}; 
Kotilainen et al. \cite{kotilainen}; Kotilainen \& Ward \cite{kotilainen-ward}).
In particular, galactic bars which should facilitate the inward transport of
gas to fuel the active nucleus (Shlosman, Begelman, \& Frank
\cite{shlosman:nature}) are not ubiquitous in Seyfert galaxies, even 
at the infrared wavelengths thought to favor bar detection
(McLeod \& Rieke \cite{mcleod}; Mulchaey \& Regan \cite{mulchaey}).
Likewise, interactions are frequent in Seyferts
(Dahari \cite{dahari}; MacKenty \cite{mackenty89}), but not all Seyferts are
found in interacting systems (De Robertis, Yee, \& Hayhoe \cite{derobertis}).
The only salient difference between Seyfert and normal spiral {\it morphology}
discovered to date is the near-infrared surface brightness of the disk 
(Hunt et al. \cite{hunt2}): at 2\,$\mu$m, Seyfert disks turn 
out to be almost 1\,mag\,arcsec$^{-2}$ brighter than those in normal
early-type spirals. 

This paper is the first of a series aimed at the investigation
of the host-galaxy/nuclear connection in AGNs and starbursts
through qualitative and then quantitative galaxy morphology.
Our study is based on the Extended 12\,$\mu$m Galaxy Sample
(Rush, Malkan, \& Spinoglio \cite{rms} -- hereafter RMS), and
here we report an analysis, with data from the literature,
of the Hubble types, axial ratios, and bar and ring fractions. 
We also assess the impact of the infrared selection criterion on the star
formation properties of the sample galaxies.
Future papers will present and analyze new optical and near-infrared images of
the 12\,$\mu$m Seyferts that will enable us to further quantify the morphologies
of these objects. 

\section{The Extended 12\,$\mu$m Galaxy Sample}

We have chosen the Extended 12\,$\mu$m Galaxy Sample (hereafter E12GS) 
for several reasons.
(1)~The 12\,$\mu$m selection avoids biases associated with 
ultraviolet selection criteria which tend to favor 
blue Seyfert 1s and quasars [e.g., Markarian and Green (Green,
Schmidt, \& Liebert \cite{green})].
(2)~Optically-selected magnitude-limited samples may be
biased against faint nuclei embedded in bright galaxies
(e.g., CfA Seyfert sample: Huchra \& Burg \cite{huchra:burg}), but
the 12\,$\mu$m flux is  an approximately constant fraction of the 
bolometric flux in both types of Seyferts (RMS).
(3)~Importantly for statistical studies, the 12\,$\mu$m Seyferts
are numerous (116) and comprise the largest Seyfert sample of both types
yet compiled. 
(4)~12\,$\mu$m Seyferts are closer than the CfA Seyferts, thus
affording better spatial resolution and higher flux densities.
(5)~Distances of the two types 
in the 12\,$\mu$m Seyferts are similar (69~Mpc for Seyfert 1s vs. 59~Mpc
for Seyfert 2s) so that
conclusions drawn from type comparison should not suffer
from resolution or distance/luminosity effects.
(6)~Enhanced star formation activity may be favored by the infrared
selection criterion, and evaluation of such a selection artifact may
help better understand the relationship between Seyfert activity and
star formation. 
(7)~Finally, the E12GS automatically guarantees similarly-selected control samples
of HII/starbursts, LINERS, and non-active galaxies.

The E12GS was defined by RMS
on the basis of 12\,$\mu$m flux, and contains 891 
galaxies\footnote{3C273 and OJ287 are not considered here.}.
The flux limit is 0.22~Jy, and the sample is estimated to be complete to 0.3~Jy.
The multiwavelength properties of Seyferts in the 12\,$\mu$m sample are discussed
by Rush \& Malkan (\cite{rush1}); Rush et al. (\cite{rush2}); and 
Rush, Malkan, \& Edelson (\cite{rush3}).
Near-infrared photometry of the galaxies in general is reported by
Spinoglio et al. (\cite{spinoglio}).

\section{Morphology of the 12\,$\mu$m Galaxies}\label{sample}\protect

We are interested in the morphological characteristics of the 12\,$\mu$m galaxies
not only for their intrinsic interest,
but also to assess any dependence on the 12\,$\mu$m selection criterion. 
Specifically,
{\it (i)}~what are the Hubble types 
of the 12\,$\mu$m galaxies, and how do they compare with those for normal spirals
and previous Seyfert samples?
{\it (ii)}~what are the axial ratios of the 12\,$\mu$m galaxies,
and how do they compare with optically-selected samples?
{\it (iii)}~how does the bar fraction of the 12\,$\mu$m galaxies compare with that
for normal spirals, and how does it vary with activity class and Hubble
type?
{\it (iv)}~what fraction of 12\,$\mu$m galaxies have rings?

To this end, we used the spectroscopic 
classifications of the E12GS  
(HII/starburst, LINER, Seyfert 1 or 2)
from the NASA/IPAC Extraglactic Database,
NED\footnote{The NASA/IPAC Extragalactic Database (NED)
is operated by the Jet Propulsion Laboratory, 
California Institute of Technology, under contract with the U.S.
National Aeronautics and Space Administration.}.
From these classifications, three subsets of the 
12\,$\mu$m galaxies were derived: HII/starburst (67 objects), 
LINER (33 objects), and ``non-active'' (neither HII/starbursts, nor LINERs, nor Seyferts:
626 objects, hereafter referred to as ``normal''). 
Ambiguous designations such as HII$+$LINER (8 objects), HII$+$Sy (6), LINER$+$Sy (14)
have been separated out in order to better represent ``pure'' activity
classes, although we have analyzed them where necessary to evaluate
the validity of possibly low-significance results.

The resulting percentage of ``active'' galaxies in the E12GS is roughly 30\%,
as opposed to the 20\% value given in RMS.
The increase is in large part due to the different definition and consequent
greater number of HII/starbursts here, and to the inclusion of ambiguous 
classifications.
More observations in the literature and continuous updates of NED also contribute
to the increase.
29 galaxies are classified as Seyferts in NED, but are not part
of the 12\,$\mu$m Seyfert samples, and, conversely, 22 of the 116 
12\,$\mu$m Seyferts are not designated as such in NED.
Hence, there may be some doubt about the strict membership of the 
activity-class subsamples we have defined, although the samples should be large 
enough to submerge small random effects of mistyping.

Morphological types, bar and ring classes, 
and major and minor diameters were taken from NED,
and the distributions of these compared among the subsets defined above. 
The optically-selected CfA sample of Seyferts (Huchra \& Burg \cite{huchra:burg}) 
is also considered in the analysis, so as to better assess any selection effects 
introduced by the 12\,$\mu$m criterion (see $\S$~\ref{sfm}).

\subsection{Morphological Types }

The NED revised morphological types are typically taken from the Third
Reference Catalogue of Bright Galaxies (RC3; de Vaucouleurs et al. \cite{rc3}).
The Hubble stage or type index (T) is derived from these according to the 
principles outlined in RC3.
When an object is tagged ``pec''(uliar), but has a well-defined type, we 
included it in the analysis.
``Peculiar'' morphology is shown by only 19\% of the normal 12\,$\mu$m galaxies,
by roughly 25\% of the 12\,$\mu$m Seyferts (21 and 27\% for Types 1 and 2, 
respectively), by 35\% of the LINERs, and
by almost half (45\%) of the 12\,$\mu$m HII spirals. 

\begin{figure}
\centerline{\epsfig{figure=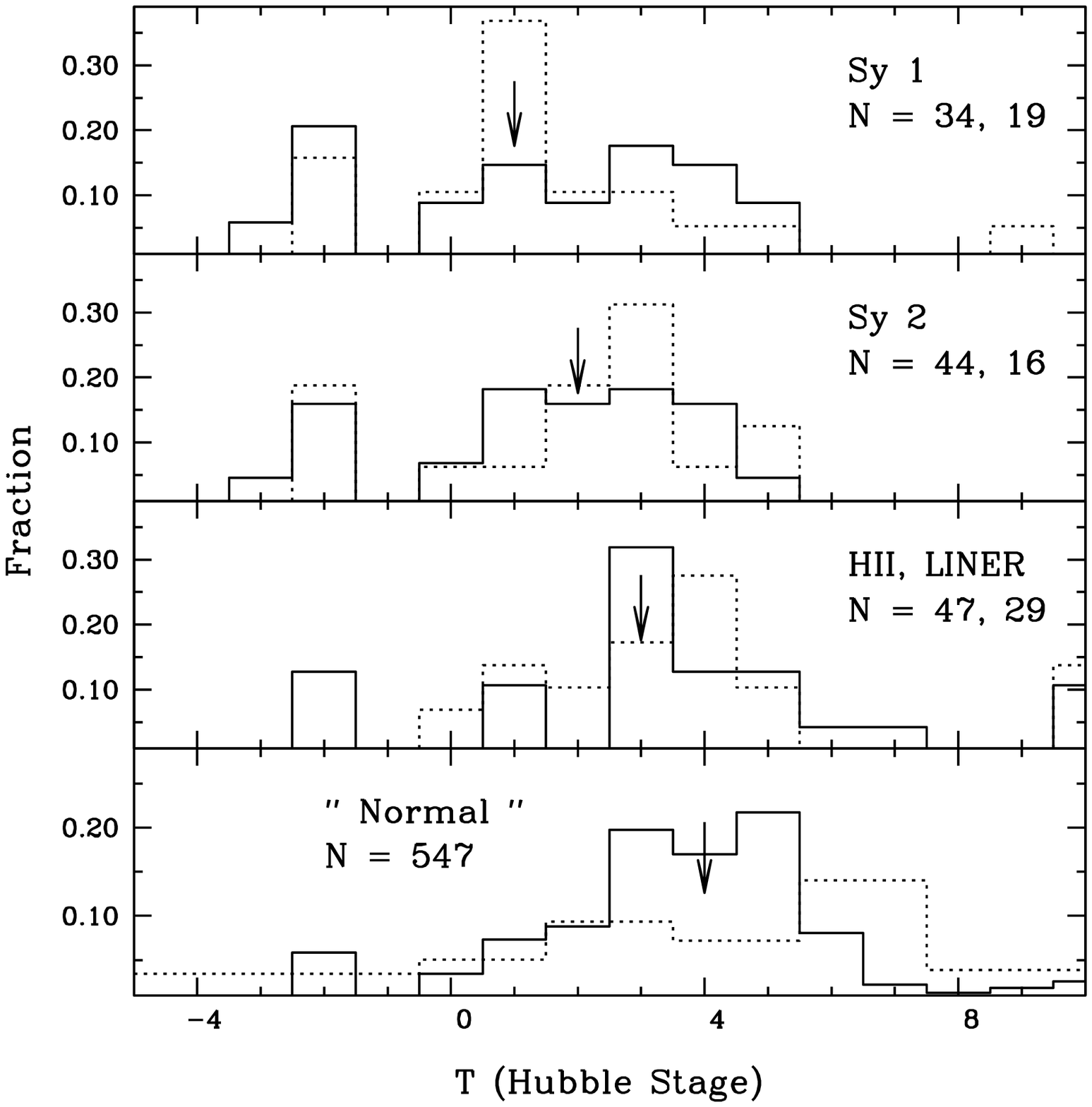,width=10cm}}
\figcaption[morph.ps]{Distributions of the morphological type index (Hubble stage). 
The distributions denoted by a dotted line correspond 
(from bottom to top) to:
1)~the distribution of
the Uppsala General Catalog as tabulated by Roberts \& Haynes 
(\cite{roberts:haynes});
2)~the 12\,$\mu$m LINER sample;
3)~the CfA Seyfert 2s;
4)~the CfA Seyfert 1s.
The numbers given in this and subsequent histograms refer to the number
of objects with (in this case) morphological types defined in NED.
The two values given in the HII/LINER panel correspond to the HII and
LINER subsamples, respectively, and
the two values in the Seyfert panels to the 12\,$\mu$m and
CfA samples.
The vertical arrows mark subsample medians,
calculated with a type index resolution of unity.
\label{fig:morph}
} 
\end{figure}

The distributions of Hubble type index for the different subsamples
are shown in Fig. \ref{fig:morph}.
It is apparent that the normal 12\,$\mu$m galaxies are predominantly spirals,
as expected, and that there are fewer very late-type spirals relative
to the UGC distribution (although both have median T~=~4~=~Sbc).
The 12\,$\mu$m and CfA Seyfert 1s tend to be early-type spirals (median T~=~1~=~Sa), 
while the 12\,$\mu$m HII galaxies and LINERs tend towards later types (medians 
T~=~3,\,3.5~=~Sb,\,Sbc, respectively); 
Seyfert 2s are intermediate between the two with median T~=~2~=~Sab.

That Seyferts tend to reside in early-type spirals has been known for some time
(Terlevich, Melnick, \& Moles \cite{tmm};
Moles, M\'{a}rquez, \& P\'erez \cite{moles}).
The trend found here of morphological types in Seyfert galaxies
is also similar to that of the Palomar Spectroscopic Survey (Ho, Filippenko, \& Sargent 
\cite{ho:iii}), and of a large sample of nearby Seyferts
(Malkan, Gorjian, \& Tam (\cite{malkan}; hereafter MGT). 
However, the 12\,$\mu$m HII galaxies tend towards earlier types and the LINERs
towards later types than those detected in the Palomar survey
(Ho, Filippenko, \& Sargent \cite{ho:v}).
This may be a luminosity effect since the Palomar Survey tends towards low luminosities,
or it could be related to the 12\,$\mu$m selection and dust or gas content.
The median morphological type of both
12\,$\mu$m and CfA  Seyfert 2s lies between  Seyfert 1s and HII galaxies/LINERs
(individual typing uncertainties are T~$\sim$~0.7--Buta et al. \cite{buta:1994}), 
a trend confirmed by the ambiguous sub-samples: LINER/Sy galaxies have 
median T~=~2, HII/Sys T~=~3, and HII/LINERs T~=~4.
There appears to be a global progression from normal spirals and LINERS
(median Sbc), to HII/starbursts (Sb), to Seyfert 2s (Sab), and Seyfert 1s (Sa).

\subsection{Axial Ratios}

\begin{figure}
\centerline{\epsfig{figure=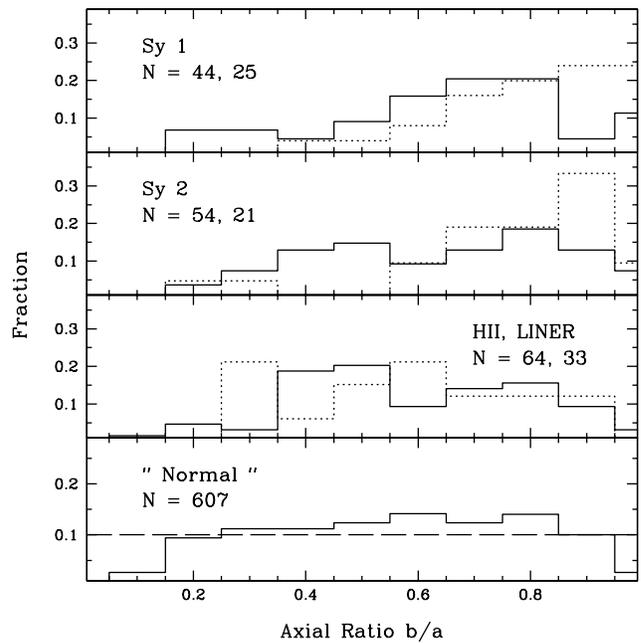,width=10cm}}
\figcaption[incl.ps]{Distributions of the axial ratios ($b/a$).
The distributions denoted by a dotted line are the same as those in
Fig. \ref{fig:morph}.
Numbers under the panel label give the number of galaxies in each subsample
with defined axial ratios.
The horizontal dashed line in the lowest panel indicates a random
distribution in the absence of optical depth effects,
and assuming disks are intrinsically round.
\label{fig:incl}
} 
\end{figure}

The intrinsic shape of spiral disks may be derived from distributions
of axial ratios and has been extensively studied
(Sandage, Freeman, \& Stokes \cite{sfs};
Binney \& de Vaucouleurs \cite{binney:dev};
Fasano et al. \cite{fasano};
Odewahn, Burstein, \& Windhorst \cite{odewahn}).
Here we compare the axial ratios in the E12GS with
those of normal spirals, and with optically-selected Seyfert samples.
The distributions of the axial ratios of the galaxies in the E12GS
are shown in Fig. \ref{fig:incl}.
The normal 12\,$\mu$m galaxies exhibit an axial-ratio distribution
entirely consistent with normal disk galaxies, being roughly flat
from $b/a\ \lesssim\ 1.0$ down to $\sim$\ 0.2
(Binney \& de Vaucouleurs \cite{binney:dev}).

Optically-selected Seyfert galaxies tend to avoid edge-on
systems (Keel \cite{keel:1980}), partly because of the bias
of (optical) magnitude-limited samples against extremely inclined
systems (Burstein, Haynes, \& Faber \cite{bhf}; Fasano et al. \cite{fasano};
Maiolino \& Rieke \cite{mr}).
Indeed, the CfA Seyferts show a strong deficiency of highly inclined
disks, especially for the type 1s:  the mean $b/a$ for the CfA type 1s (2s) is 
0.80 (0.77), compared with 0.56 for the normal 12\,$\mu$m galaxies.
Figure \ref{fig:incl} shows that
the 12\,$\mu$m Seyfert galaxies are much less affected by this
bias than the CfA sample, as 
the mean $b/a$ for the 12\,$\mu$m type 1s (2s) is 0.65 (0.64). 
11\% of the type 2 Seyferts, and 13\% of the type 1s have $b/a$\,$\leq$\,0.4,
versus none of the CfA Seyferts. 
A similar lack of edge-on Seyfert 2s has been noted by McLeod \& Rieke (\cite{mcleod}),
who attributed it to obscuration on scales of $\sim$\,100~pc or more.

The LINERs and HII galaxies have a mean $b/a$ of 0.59.
Only 9\% of the HII galaxies have $b/a$\,$\leq$\,0.4, which implies that 
if obscuration causes the deficiency of highly inclined galaxies, it
operates equally well whether an AGN or a starburst is present.

\subsection{Bars}

70\% of the normal sample and
roughly half of each of the active samples have bar classes defined,
and those galaxies with bar definitions have been divided into three categories:
unbarred (SA); barred (SAB or SB); and strongly barred (SB). 
Table \ref{tbl:bars} gives the results of this division.
It turns out that 69\% of the normal galaxies are barred, in good agreement
with the magnitude-limited sample of northern spirals with $B\,\leq\,$13
of which 68\% have bars (see Moles et al. \cite{moles}),
but slightly higher than the 60\% fraction in field spirals
(Sellwood \& Wilkinson \cite{sellwood})
and in the Palomar Survey spirals (Ho, Filippenko, \& Sargent \cite{ho:bars}).
If we apply a velocity constraint and consider only those sources with
$v\,<\,$5000\,km\,s$^{-1}$, the bar fraction is 68\%.
We conclude that the bar properties of the 12\,$\mu$m normal galaxies
are similar to those of optically-selected samples
(see also Pompea \& Rieke \cite{pompea}). 

\begin{deluxetable}{lrrrrr}
\tablecolumns{6}
\tableheadfrac{0.1}
\tablewidth{0pt}
\tablenum{1}
\tablecaption{Bar Class and Activity Type \label{tbl:bars}}
\tablehead{
& & &\multicolumn{1}{c}{Weakly} & \multicolumn{1}{c}{Strongly} \\ 
\multicolumn{1}{c}{Activity} & & \multicolumn{1}{c}{Unbarred} 
& \multicolumn{1}{c}{Barred} & \multicolumn{1}{c}{Barred} & \multicolumn{1}{c}{Barred} \\
\colhead{Type} & \colhead{Total\tablenotemark{a}} & \colhead{(SA)} & \colhead{(SAB)} & \colhead{(SB)} 
&  \colhead{(SAB\,$+$\,SB)} }
\startdata
Normal         & 447 & 138 (31\%) & 131 (29\%) & 178 (40\%) & 309 (69\%) \nl 
HII/starburst  &  34 &   5 (15\%) &  11 (32\%) &  18 (53\%) &  29 (85\%) \nl 
LINER          &  23 &   9 (39\%) &   9 (39\%) &   5 (22\%) &  14 (61\%) \nl 
12\,$\mu$m Sy 1 &  28 &   9 (32\%) &  12 (43\%) &   7 (25\%) &  19 (68\%) \nl
12\,$\mu$m Sy 2 &  33 &  11 (33\%) &   9 (27\%) &  13 (39\%) &  22 (67\%) \nl
CfA Sy 1       &  18 &   8 (44\%) &   5 (28\%) &   5 (28\%) &  10 (56\%) \nl 
CfA Sy 2       &  12 &   4 (33\%) &   5 (42\%) &   3 (25\%) &   8 (67\%) \nl 
\tablevspace{12pt}
\multicolumn{2}{c}{\bf $v\,<\,$5000 km\,s$^{-1}\quad\Rightarrow$} \nl
\tablevspace{7pt}
Normal         & 401 & 128 (32\%) & 122 (30\%) & 151 (38\%) & 273 (68\%) \nl 
HII/starburst  &  28 &   5 (18\%) &  10 (36\%) &  13 (46\%) &  23 (82\%) \nl 
LINER          &  23 &   9 (39\%) &   9 (39\%) &   5 (22\%) &  14 (61\%) \nl 
12\,$\mu$m Sy 1 &  20 &   6 (30\%) &   9 (45\%) &   5 (25\%) &  14 (70\%) \nl
12\,$\mu$m Sy 2 &  27 &  10 (37\%) &   6 (22\%) &  11 (41\%) &  17 (63\%) \nl
CfA Sy 1       &  11 &   4 (36\%) &   5 (45\%) &   2 (18\%) &   7 (64\%) \nl 
CfA Sy 2       &   6 &   2 (33\%) &   2 (33\%) &   2 (33\%) &   4 (67\%) \nl 
\enddata
\tablenotetext{a}{Total number of objects with well-defined bar class.} 
\end{deluxetable}

In contrast to the normal spirals, 
the vast majority of the 12\,$\mu$m starbursts are barred. 
82\,--\,85\% (depending on whether the distant galaxies are excluded) 
of the HII/starbursts with known bar class are barred, and half of them
are strongly 
barred\footnote{The hybrid HII/Sy sample with defined bar class has 4 of 4, or 100\%, barred,
while the HII/LINERs have 4 of 7 barred (57\%).}.
This is a significantly higher barred fraction of HII galaxies
than the 61\% found by Ho et al. (\cite{ho:bars}), and it does not
appear to be a selection effect.
A detailed discussion of barred starbursts or star formation
in barred galaxies is beyond the scope of this paper,
but it would appear that all starbursts, at least relatively luminous ones, 
are preferentially barred.
We have analyzed the bar fraction of the Markarian nuclear starbursts 
(as listed by Balzano \cite{balzano} and by 
Mazzarella \& Balzano \cite{mazzarella}) and found that 87.5\% 
of the (32) galaxies with known bar class are barred, and 
75\% are strongly barred.

The bar fractions of 12\,$\mu$m LINERs and Seyferts appear very normal with 
between 61 and 68\% of them having bars; the CfA Seyferts are similar
with roughly a 62\% bar fraction.
4 of 9 hybrid LINER/Sy are barred (56\%).
This is yet another confirmation of the emerging consensus that bars
occur in Seyferts with the same frequency as they occur in the normal spiral 
population (see Introduction and references therein).
Recent work has suggested that bars and distortions are more frequent
in type 2 Seyferts than in type 1s (Maiolino et al. \cite{maiolino}).
For the 12\,$\mu$m Seyferts, however, this is not the case as the bar
fraction (with $v\,<$\ 5000\,km\,s$^{-1}$) is greater
for type 1s (70\%) than for type 2s (63\%). 
It is only marginally the case for the CfA sample which has
67\% of Seyfert 2s barred versus 64\% of Seyfert 1s, although if
all distances are considered, 
the differences between the two types in the CfA sample are more pronounced 
(56\% type 1s vs. 67\% type 2s, see Table \ref{tbl:bars}).
More and higher-quality image data are needed to
decide if the difference is significant, and if so, if it results
from optical selection.

\begin{figure}
\centerline{\epsfig{figure=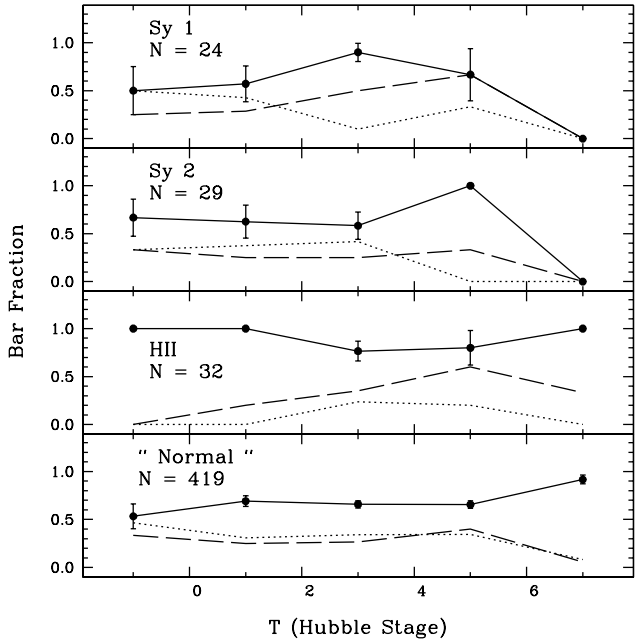,width=10cm}}
\figcaption[bar2.ps]{Fraction of barred galaxies as a function of Hubble type. 
The dotted lines shows the unbarred galaxies, the dashed lines the
weakly-barred (SAB), and the solid lines the barred (SAB\,$+$\,SB).
Error bars are shown only for the barred distributions,
and are derived from counting statistics as 
$\sqrt{f(1-f)/N}$, where $f$ is the fraction of barred objects in a given
morphological type bin, and $N$ is the total number of objects in the bin.
Numbers under the panel label give the number of galaxies in each subsample 
with well-defined bar class.
The data have been binned as follows:
S0a and earlier (T\,$\leq$\,0); 
Sa, Sab (0\,$<$\,T\,$\leq$\,2);
Sb, Sbc (2\,$<$\,T\,$\leq$\,4);
Sc, Scd (4\,$<$\,T\,$\leq$\,6);
Sd and later (6\,$<$\,T).
\label{fig:bar2}
}
\end{figure}

We turn finally to the variation of bar fraction with morphological type.
Although the counting statistics are sufficiently large only for the
normal sample, we have analyzed bar fraction as a function
of Hubble type as shown in Fig. \ref{fig:bar2};
the data have been binned as described in the figure caption.
Very late-type normal E12GS galaxies show a higher percentage of bars
than early types, as also found by Ho et al. (\cite{ho:bars}):
86\% (31/36) of the 12\,$\mu$m galaxies with  T~$>$~6 are {\it strongly} 
barred (SB), and only 8\% (3/36) are not barred at all.
Otherwise the bar fraction is constant with morphological type, except
for S0s and earlier, which tend to have a lower percentage
(see also Ho et al.).

Within the errors, the bar fractions of 12\,$\mu$m HII galaxies and Seyfert 2s 
are also constant with morphological type, while Seyfert 1s show a peak of 90\%
at type Sb (T~=~3).
This type is also the mode in the Hubble type distribution, and means that
9 of 10 Sb Seyfert 1s are barred.
The fraction of unbarred Seyfert 1s is highest for the very early-types
(S0 and earlier), similar to the case for normal galaxies.

\subsection{Rings}

\begin{figure}
\centerline{\epsfig{figure=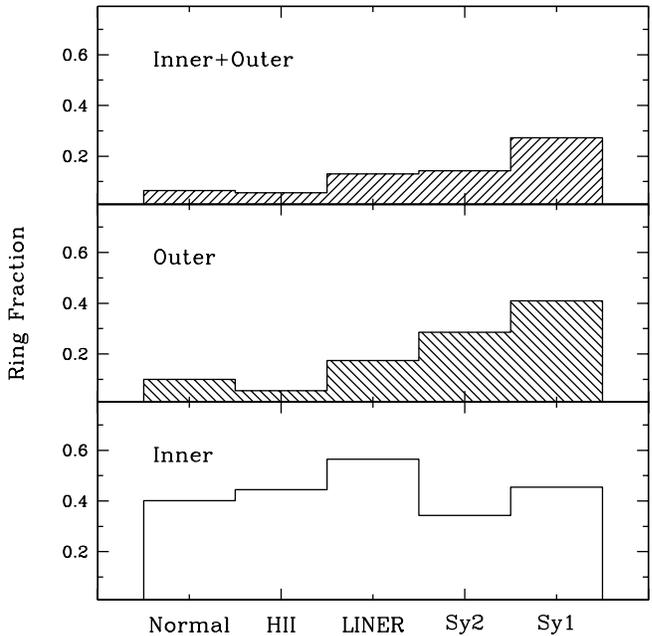,width=10cm}}
\figcaption[rings.ps]{Fraction of ringed galaxies as a function of
activity class.
The lower panel shows inner rings, the middle panel outer ones, and
the upper panel those galaxies with both inner and outer rings.
Only those objects with ($v\,<$\,5000\,km\,s$^{-1}$) are shown in
the figure.
\label{fig:rings}
}
\end{figure}

Rings trace dynamical resonances in galaxies and are locations
of strong density enhancements in stars and gas.
Numerical simulations show that bar perturbations can form rings (Schwarz \cite{schwarz}),
and almost inevitably do in early-type spirals 
(Combes \& Elmegreen \cite{combes:elmegreen}; Piner et al. \cite{piner}).
Patterns in ring structure have also been interpreted as an evolutionary sequence
in Seyferts where rings signify active gas flow into nuclear regions
(Su \& Simkin \cite{ss}).
For these reasons, we have quantified the presence of rings in the E12GS.
Inner and outer rings have been associated with inner and outer Lindblad resonances
(Schwarz \cite{schwarz}; Buta \cite{buta:1993}; Piner et al. \cite{piner}), and we 
have used the RC3 designations ``{\it (r)}'' and ``{\it (rs)}'' to tally inner rings, 
and ``{\it (R)}'' and ``{\it (R')}'' for outer ones.
The tallies are shown in Table \ref{tbl:rings}, together with the $(s)$ (S shape),
which designates an inner spiral.
Following Simkin et al. (\cite{sss}), we have applied a velocity constraint
($v\,<$\,5000\,km\,s$^{-1}$), shown in the lower part of the Table and in
Fig. \ref{fig:rings}, so as to ensure
reasonably consistent typing among the different activity types.

Outer rings are relatively rare in normal galaxies, as shown in
Table \ref{tbl:rings}. 
Simkin et al. (\cite{sss}) found similar fractions for an
optically-selected comparison sample of normal spirals, so
the 12\,$\mu$m galaxies appear normal in terms of the occurrence of rings.
Three features, however, stand out in Table \ref{tbl:rings}:
(1)~the high fraction of Seyferts -- especially type 1s --
with outer rings (40\% for $v\,<$\,5000\,km\,s$^{-1}$,
as opposed to 10\% for the normal 12\,$\mu$m galaxies);
(2)~the high fraction of Seyferts and LINERs with both inner and outer rings;
and (3)~the high fraction of LINERs with inner rings.
Seyferts show outer rings or simultaneous inner and outer ring structures 
between 3 and 4 times more frequently than do normal spirals;
the formal significance of the higher Seyfert 1 outer ring fraction is 3.9\,$\sigma$.
Outer rings occur in LINERS almost twice as often as normal, although with low
significance, and with a normal frequency in HII/starburst galaxies. 
No outer rings are found in any of the hybrid activity classes,
although S shapes occur in 70\% of the HII hybrids (HII/LINER, HII/Sy).
The anomalously high outer ring fraction in Seyferts
is similar to results reported for Simkin et al. (\cite{sss}) for a smaller 
optically-selected sample.
The incidence of inner rings (40\%) and S shapes (33\%) in normal spirals 
(see also Simkin et al. \cite{sss}) is exceeded only by the LINERs (57\%), 
a 2-$\sigma$ effect.
MGT also found that inner rings are not unusually frequent in Seyfert galaxies.

\begin{deluxetable}{lrrrrrr}
\tablecolumns{7}
\tableheadfrac{0.1}
\tablewidth{0pt}
\tablenum{2}
\tablecaption{Ring Class and Activity Type \label{tbl:rings}}
\tablehead{
&& \multicolumn{1}{c}{Outer} & \multicolumn{1}{c}{Inner} \\
\multicolumn{1}{c}{Activity} && \multicolumn{1}{c}{Ring} & \multicolumn{1}{c}{Ring} 
& \multicolumn{1}{c}{Outer$\,+\,$} 
& \multicolumn{1}{c}{Outer$\,||\,$} 
& \multicolumn{1}{c}{S-shaped} \\
\colhead{Type} & \colhead{Total\tablenotemark{a}}
& \colhead{$(R)\,+\,(R^\prime)$} & \colhead{$(r)\,+\,(rs)$} & \colhead{Inner\tablenotemark{b}} 
& \colhead{Inner\tablenotemark{c}} 
& \colhead{$(s)$} }
\startdata
Normal          & 547 &  58 (11\%) & 209 (38\%) &  36 (\phn 7\%) 
& 237 (43\%) & 174 (32\%) \nl 
HII/starburst   &  47 &   2 (\phn 4\%) &  18 (38\%) &   2 (\phn 4\%) 
& 18 (38\%) &  13 (28\%)\nl 
LINER           &  29 &   4 (14\%) &  13 (45\%) &   3 (10\%) 
& 14 (48\%) &   8 (28\%)\nl 
12\,$\mu$m Sy 1 &  34 &  10 (29\%) &  13 (38\%) &   6 (18\%) 
& 17 (50\%) &  12 (35\%)\nl 
12\,$\mu$m Sy 2 &  45 &  11 (24\%) &  14 (31\%) &   5 (11\%) 
& 21 (47\%) &  11 (24\%)\nl 
CfA Sy 1        &  19 &   6 (32\%) &   6 (32\%) &   2 (11\%) 
& 10 (53\%) &  10 (53\%)\nl 
CfA Sy 2        &  16 &   3 (19\%) &   7 (44\%) &   3 (19\%) 
& 7 (44\%) &   1 (\phn 6\%) \nl 
\tablevspace{12pt}
\multicolumn{2}{c}{\bf $v\,<\,$5000 km\,s$^{-1}\quad\Rightarrow$} \nl
\tablevspace{7pt}
Normal          & 479 & 48 (10\%)    & 192 (40\%) & 31 (\phn 6\%) 
& 215 (45\%) & 162 (34\%) \nl
HII/starburst   &  36 &  2 (\phn 6\%) &  16 (44\%) &   2 (\phn 6\%) 
& 16 (44\%) &  11 (31\%)\nl  
LINER           &  23 &   4 (17\%) &  13 (57\%) &   3 (13\%) 
& 14 (61\%) &   8 (35\%)\nl  
12\,$\mu$m Sy 1 &  22 &   9 (41\%) &  10 (45\%) &   6 (27\%) 
& 13 (59\%) &   9 (41\%)\nl 
12\,$\mu$m Sy 2 &  35 &  10 (29\%) &  12 (34\%) &   5 (14\%) 
& 18 (51\%) &   9 (26\%)\nl 
CfA Sy 1        &  10 &   4 (40\%) &   4 (40\%) &   2 (20\%) 
& 6  (60\%) &   7 (70\%)\nl 
CfA Sy 2        &   8 &   2 (25\%) &   3 (38\%) &   2 (25\%) 
& 3  (38\%) &   1 (12\%)\nl  
\enddata
\tablenotetext{a}{Number of objects with well-defined morphological type.} 
\tablenotetext{b}{Number of objects with both an inner and an outer ring.} 
\tablenotetext{c}{Number of objects with either an inner or an outer ring.} 

\end{deluxetable}

Because inner and outer rings may occur preferentially in very early-type barred
spirals (Combes \& Elmegreen \cite{combes:elmegreen}; Elmegreen et al. \cite{eecb}), 
we have checked to ensure that the high outer ring fraction in Seyferts is not due
to their predominantly early Hubble types.
Considering only 12\,$\mu$m normal galaxies with T$\,\leq\,$2 (Sab), 
the outer ring fraction increases by a factor of two (to 24\%),
while the inner ring fraction remains roughly constant (at 43\%).
Among only early types,
the Seyfert outer ring fractions are 60\% and 50\%, for
Types 1 and 2, respectively. 
These fractions remain significantly higher than normal,
although instead of a 4-$\sigma$ effect, it is around 2\,$\sigma$. 

We can also examine what percentage of barred and unbarred galaxies have rings,
as shown in Table \ref{tbl:barsrings}.
While 10\% of all normal 12\,$\mu$m galaxies have outer rings, 13\% of
barred normal spirals do, and 7\% of the unbarred spirals.
While the presence of a bar may influence the occurrence of
an outer ring, bars definitely appear to be associated with inner rings:
52\% of barred spirals have inner rings, but only 
38\% of unbarred spirals have them. 
Such a result is not surprising since inner resonances are expected to develop 
in early-type barred spirals (Combes \& Elmegreen \cite{combes:elmegreen}).
These fractions for 12\,$\mu$m spirals are consistent with those reported 
by de Vaucouleurs \& Buta (\cite{buta:1980}) who find a fraction of 43\% 
of the unbarred and almost 70\% of the barred spirals in the Second 
Reference Catalogue of Bright Galaxies 
(RC2; de Vaucouleurs, de Vaucouleurs, \& Corwin \cite{rc2}) to have inner rings.
Finally, twice the normal fraction ($\gtrsim$\,80\%)
of {\it unbarred} HII/starburst galaxies have inner rings.

\begin{deluxetable}{lrrrr}
\tablecolumns{5}
\tableheadfrac{0.1}
\tablewidth{0pt}
\tablenum{3}
\tablecaption{Bars with Rings and Activity Class\tablenotemark{a}\label{tbl:barsrings}}
\tablehead{
\multicolumn{1}{c}{Activity} & \multicolumn{2}{c}{$(R)\,+\,(R^\prime)$} 
& \multicolumn{2}{c}{$(r)\,+\,(rs)$} \\
\colhead{Type} & \colhead{Barred\tablenotemark{b}} & \colhead{Unbarred\tablenotemark{b}} 
& \colhead{Barred\tablenotemark{b}} & \colhead{Unbarred\tablenotemark{b}} }
\startdata
Normal          &  36 (13\%) &   9 (\phn 7\%) & 142 (52\%) &  49 (38\%) \nl 
HII/starburst   &   1 (\phn 4\%) &   1 (20\%) &  12 (52\%) &   4 (80\%) \nl 
LINER           &   2 (14\%) &   2 (22\%) &   8 (57\%) &   5 (56\%) \nl 
12\,$\mu$m Sy 1 &   8 (57\%) &   1 (17\%) &   9 (64\%) &   1 (17\%) \nl 
12\,$\mu$m Sy 2 &   8 (47\%) &   2 (20\%) &  10 (59\%) &   2 (20\%) \nl
CfA Sy 1        &   4 (57\%) &   0 (\phn 0\%) &   4 (57\%) &   0 (\phn 0\%) \nl
CfA Sy 2        &   1 (25\%) &   1 (50\%) &   2 (50\%) &   1 (50\%) \nl
\enddata
 
\tablenotetext{a}{With $v\,<\,5000$~km\,s$^{-1}$.} 
\tablenotetext{b}{Percentages are calculated using total number with a given 
bar class.  (See Cols. (3) and (6) of Table \ref{tbl:bars}.)} 

\end{deluxetable}

\subsection{Morphology Summary}

We have examined the attributes, gleaned from NED, of the E12GS, and find: 
\renewcommand{\theenumi}{{\it \roman{enumi}}}
\begin{enumerate}
\item
The normal spirals in the E12GS are truly normal in terms of 
morphological types, axial ratio distribution, bar fraction, and rings.
Only one-fifth (19\%) of them are designated as morphologically ``peculiar''. 
\item
Almost half (45\%) of the 12\,$\mu$m HII/starburst galaxies are morphologically
``peculiar'', and more than 80\% of them are barred. 
A normal percentage of them has rings, but twice the normal fraction 
($\gtrsim$\,80\%) of unbarred HII/starbursts show inner rings.
\item
The axial ratios of the 12\,$\mu$m Seyferts show
a much smaller deficiency of edge-on systems
than in optically-selected Seyferts.
This is because selection in the mid-infrared is much less biased
against Seyferts suffering large extinction.
\item
One-fourth of the 12\,$\mu$m Seyferts are classified as having ``peculiar''
morphology, and, consistently with previous studies, they are not
preferentially barred.
\item
The incidence of outer rings and simultaneous inner/outer rings
in the 12\,$\mu$m Seyferts is overwhelmingly higher ($\times\,$3--4)
than that for normal spirals; 
LINERs show a high ($\times\,$1.5 normal) inner ring fraction.
Both results are statistically significant.
\end{enumerate}

\section{Star Formation in the 12\,$\mu$m Sample}\label{sfm}\protect

An important selection effect that may operate in the E12GS
is the tendency for infrared-selected galaxies
to be dominated by powerful star formation (e.g., Soifer et al. \cite{soifer1}).
Star formation occurring on short timescales can alter
morphology, and thereby flavor results drawn from a morphological analysis. 
Such a selection artifact could also
influence conclusions about star formation and Seyfert activity. 
For these reasons, we have 
attempted to quantify star formation in the E12GS on the basis of 
infrared luminosity ratios and colors.

\subsection{Infrared-to-Blue Luminosity Ratio}

Infrared (IR) emission measured by the {\it Infrared Astronomical Satellite}
(IRAS) is usually attributed to dust heated
by the quiescent interstellar medium (20--25~K), young massive stars
($\sim$\,50~K), and possibly an AGN 
(e.g., Rowan-Robinson \& Crawford \cite{rrc}).
The relative contribution of these processes determines the infrared luminosity 
output which can vary substantially from galaxy to galaxy.
As the contrasting examples of M~31 and Arp~220 illustrate 
(Telesco \cite{telesco}),
infrared-to-blue luminosity ratios ($L_{IR}/L_B$) range over
a factor of 1000 (e.g., Soifer et al. \cite{soifer:mini}).
Although $L_{IR}/L_B$ is not a direct measure of ``infrared activity''
in galaxies (Soifer et al. \cite{soifer2}), it is commonly used as an indicator
of the relative importance of young stars and vigorous star formation
(Keel \cite{keel:1993}; Combes et al. \cite{combes:prugniel}),
and we have calculated $L_{IR}/L_B$ for the sample galaxies.
For the far-infared (FIR) contribution (from 40 to 120\,$\mu$m), 
we used the usual expression:
${\it FIR}~=~3.25\,\times\,10^{-14}\,f_\nu(60\,\mu m) +
1.26\,\times\,10^{-14}\,f_\nu(100\,\mu m)$
(e.g., Persson \& Helou \cite{persson}).
The $B$-band contribution was calculated as $\nu\,f_\nu$ based on the
magnitudes given in NED, taken mostly from RC3.

\begin{figure}
\centerline{\epsfig{figure=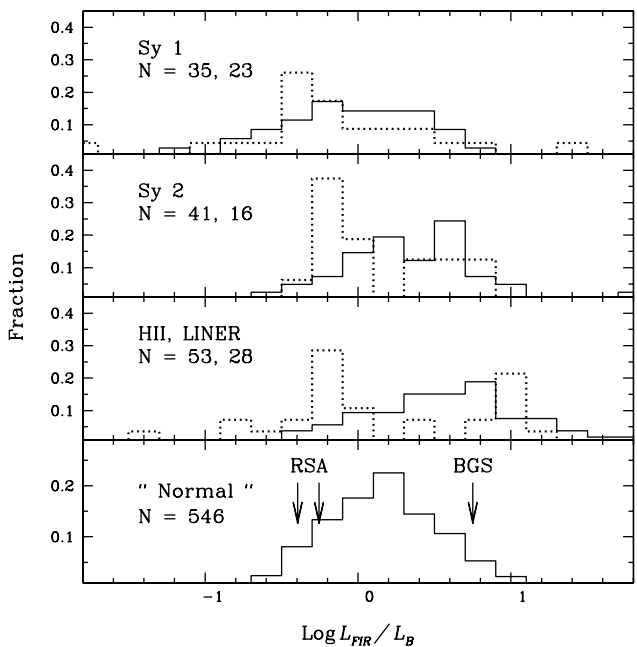,width=10cm}}
\figcaption[his.firb.ps]{Distributions of log\,$L_{IR}/L_B$. 
The distributions denoted by a dotted line are the same as those in
Fig. \ref{fig:morph}.
The vertical arrows in the lowest panel mark medians of the
unbarred (-0.4) and barred (-0.26) Shapley-Ames galaxies 
as analyzed by De Jong et al. (\cite{dejong:rsa}), and
the IRAS Bright Galaxy Sample (Soifer et al. \cite{soifer1})
selected at 60\,$\mu$m (0.75).
\label{fig:firb}
}
\end{figure}

Figure \ref{fig:firb} shows the distributions of $L_{IR}/L_B$ for the
various activity classes of the E12GS.
The normal galaxies in the E12GS are characterized by $L_{IR}/L_B$ (median
log of 0.16) intermediate
between the high ratios typical of infrared-selected galaxies 
selected at 60\,$\mu$m (IRAS Bright Galaxy Sample --BGS-- Soifer et al. \cite{soifer1};
the median (log) of 0.75 is
shown by the right vertical arrow in the lowest panel of Fig. \ref{fig:firb}),
and the low ratios in optically-selected galaxies 
[de Jong et al. \cite{dejong:rsa}, medians (log) of -0.4 (unbarred)
and -0.26 (barred) shown by the left arrows].
We conclude that
the tendency of infrared selection criteria to favor high $L_{IR}/L_B$
is much less pronounced at 12\,$\mu$m than at 60\,$\mu$m.

The 12\,$\mu$m-selected Seyferts have higher $L_{IR}/L_B$ 
(median log: -0.05, 0.31 for type 1s and 2s, respectively) than 
their optically-selected CfA counterparts ( -0.14, 0.13 for the CfA Sy 1s, 2s);
Seyfert 2s in both samples show higher values of $L_{IR}/L_B$
than in Seyfert 1s.
The 12\,$\mu$m HII/starbursts have a median (log) $L_{IR}/L_B$ (0.55)
larger than any of the other classes, comparable only to the BGS, while 
12\,$\mu$m LINERS show the lowest $L_{IR}/L_B$ of any of the 12\,$\mu$m 
activity classes, similar to those in Seyfert 1s.
Although these results seem to indicate a moderate 12\,$\mu$m selection effect on Seyferts,
some fraction of the Seyfert 60\,$\mu$m flux can be due to the AGN
(Spinoglio et al. \cite{spinoglio}).
Such a contribution would boost $L_{IR}/L_B$, independently of recent star
formation history.

\subsection{The Proportion of Warm and Cold Dust Components}

\begin{figure}
\centerline{\epsfig{figure=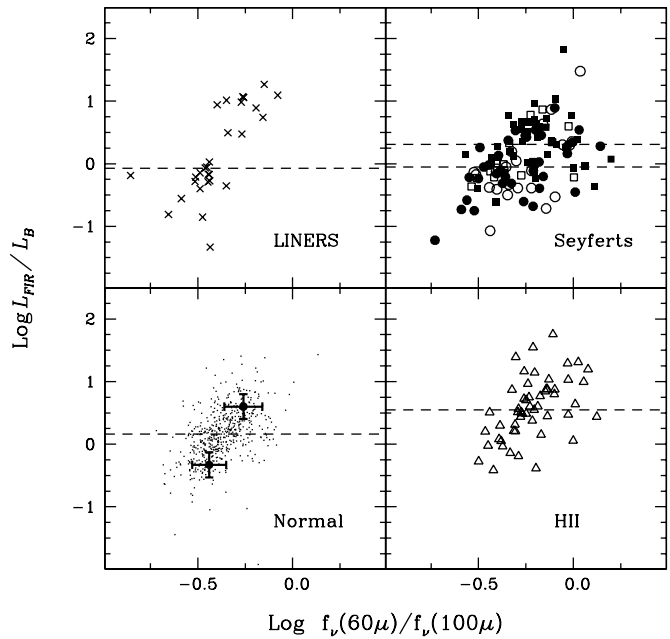,width=10cm}}
\figcaption[sfm.ps]{Plots of $L_{IR}/L_B$ versus far-infrared flux ratio
$f_\nu(60\mu)/f_\nu(100\mu)$; scale is logarithmic. 
In the upper right panel, the 12\,$\mu$m Seyferts are marked with filled 
symbols (circles for Type 1, squares for Type 2), and the CfA Seyferts
with the respective open symbols.
In the Normal panel, the two bold data points correspond to the mean and
spread in an optically-selected sample (RSA galaxies, de Jong et al.
\cite{dejong:rsa}), and in a 60\,$\mu$m-selected sample
(BGS, Soifer et al. \cite{soifer2}).
Each panel contains
dashed lines which show the median $L_{IR}/L_B$ for the activity class;
in the Seyfert panel, the upper horizontal dashed line marks Type 2s,
and the lower Type 1s.
The two outlying points (upper right) in the Seyfert panel 
are Arp~220 (12\,$\mu$m Sy2) and Mrk~231 (CfA Sy1).
\label{fig:sfm}
}
\end{figure}

IRAS observations revealed that infrared-to-blue luminosity
ratios correlate with the flux ratio $\Theta \ \equiv \ f_\nu (60\mu) /f_\nu (100\mu)$
(de Jong et al.  \cite{dejong:rsa}), although the correlation is looser 
in IR-selected samples (Soifer et al. \cite{soifer:mini}).
This was probably the first observational evidence of the
presence of a cold dust component and a more variable warm component: 
the higher the 60\,$\mu$m/100\,$\mu$m ratio, the higher the proportion of
warm dust, and the more $L_{IR}$ emitted relative to the optical.
Such correlations are shown for the E12GS in Fig. \ref{fig:sfm}.
The two bold data points in the Normal panel show the loci of optically-selected
samples (lower left, de Jong et al. \cite{dejong:rsa}), and
60\,$\mu$m-selected ones (upper right, Soifer et al. \cite{soifer2}).
It can be seen that the normal 12\,$\mu$m galaxies are well-represented by these
values, while almost half of the HII/starbursts exceed them.
On the other hand, more Seyferts fall below the normal range than
above it, and they show the worst correlation of any of the activity classes.

Subsequent work demonstrated that IR emission from the interstellar medium
(ISM), the cold component, is not
well-represented by ``classical'' dust grains in thermal equilibrium
(silicate and graphite
particles with diameters ranging from 0.005 to 0.25\,$\mu$m), as they
fail to produce the observed emission for $\lambda\,<\,40\,\mu$m
(Pajot et al. \cite{pajot}).
Very small grains transiently heated by the absorption of single UV
photons were proposed by Sellgren (\cite{sellgren}) to explain the
excess mid-infrared emission. 
In galaxies, a general relationship between IRAS color ratios
$f_\nu(60\mu)/f_\nu(100\mu)$ and $f_\nu(12\mu)/f_\nu(25\mu)$
was found by Helou (\cite{helou}), and intrepreted as 
the interplay between the contributions from classical and very small grains.
Indeed, an empirical estimate of the small-to-large dust grain ratio 
$\Gamma \ \equiv \ \nu f_{\nu} (12\mu) / FIR$ 
has been shown to depend only on the flux ratio $\Theta$ 
in Galactic nebulae and in optically- and IR-selected samples of galaxies
(Helou, Ryter, \& Soifert \cite{hrs}; hereafter HRS).

\begin{figure}
\centerline{\epsfig{figure=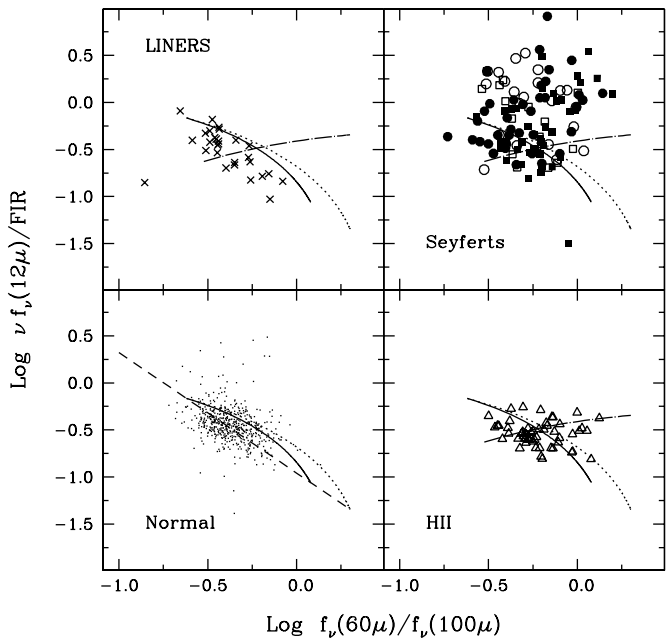,width=10cm}}
\figcaption[smg.ps]{Plots of $\Gamma\ \equiv\ \nu f_\nu (12\mu)/FIR$
versus $f_\nu(60\mu)/f_\nu(100\mu)$; scale is logarithmic.
In all the panels, the models described in HRS are shown as a solid
line and dotted line; the dashed line in the normal panel corresponds 
to the linear fit for Galactic interstellar medium data given in HRS,
and is not reproduced in the other panels. 
The monotonically-increasing dot-dashed line in the active panels
(not shown in the normal panel)
represents the trend that would be followed if
the colors were dictated by the flux limit at 12\,$\mu$m (see HRS).
The outlier (lower right) in the Seyfert panel is Arp~220.
\label{fig:smg}
}
\end{figure}

Given that the E12GS is defined on the basis of 12\,$\mu$m flux,
the selected galaxies may be anomalous in their dust content
or relative contributions from the large and small dust grains.
We have therefore considered the behavior of $\Gamma$,
the ratio of 12\,$\mu$m flux to FIR, in the sample galaxies.
Following HRS, Fig. \ref{fig:smg} shows $\Gamma$ versus 
$\Theta$ for the various activity classes in the E12GS.
The phenomenological model described in HRS, and plotted in Fig. 
\ref{fig:smg}, consists of two dust phases: a cool one represented with
a quiescent ISM energy density of 
$u_c\,\simeq\,0.5$\,eV\,cm$^{-3}$, and the other with
an increasing fraction immersed in a higher intensity field
$u_w\,\simeq\,30$\,eV\,cm$^{-3}$ (solid line), or 
$u_w\,\simeq\,100$\,eV\,cm$^{-3}$ (dotted line).
Also shown in Fig. \ref{fig:smg} as a dashed line in the lower
left panel is the linear fit to 
data from the Galactic interstellar medium in the vicinity of stars
as given by HRS. 
The remaining panels in Fig. \ref{fig:smg} show as a dot-dashed line
the colors that would be observed were the quantities artificially 
controlled by 
the 12\,$\mu$m flux limit; such a trend is not followed by the data. 
The figure shows convincingly that the HRS two-phase dust model is
appropriate for the bulk of the normal galaxies, the LINERS, and the
HII/starbursts, except, perhaps at high $\Theta$.

To assess the influence of star-formation properties on the
morphological characteristics described in the previous section,
we have calculated the fractions of bars and rings in ``star-forming''
and ``quiescent'' galaxies in the E12GS, as distinguished by the
far-infared flux ratio $\Theta$.
The median value of the HII/starbursts was used as the threshold:
$\Theta$\,=\,0.51; 
galaxies with the far-infared flux ratio greater than this value
were (arbitrarily) considered ``star-forming'', and the others ``quiescent''.
With this criterion,
20\% of the normal 12\,$\mu$m galaxies turn out to be actively star-forming,
and 50\% of the HII/starbursts (by definition), 13\% of the LINERs,
30\% of the Seyfert 1s, and 56\% of the Seyfert 2s.
The occurrence of bars in the star-forming galaxies is
significantly (4.8-$\sigma$) higher than the quiescent category 
(86\% versus 63\%) only for the normal galaxies;
there is no equivalent trend for HII/starbursts, LINERs, or Seyferts.

Normal star-forming galaxies also have a higher percentage of {\it outer} rings
than their quiescent counterparts (24\% versus 8\%), with a significance
of 2.4\,$\sigma$; the same is true for Seyfert 1s (83\% versus 28\%, 
3\,$\sigma$), but not for Seyfert 2s (40\% versus 33\%).
An opposing trend emerges for {\it inner} rings, with quiescent normal
galaxies having a higher inner ring fraction (50\%) than the
star-forming ones (37\%), an effect which is only marginally significant
at 2.1\,$\sigma$.
The same is true for Seyfert 2s (66\% quiescent versus 27\% star-forming, 
significance 2.2\,$\sigma$),
but not for Seyfert 1s. 
Evidently outer rings tend to prefer ``actively-star-forming'' galaxies, while 
inner ones prefer quiescent ones\footnote{87\% of the LINER sample is quiescent, 
and also has the highest inner ring fraction of any of the activity classes.}.
Nevertheless, even the quiescent Seyferts have outer ring fractions of
around 30\%. We conclude then that the high outer ring fraction found in 
Seyferts is not an artifact of the the 12\,$\mu$m selection criterion, which
may favor more recent star formation.

\subsection{Mid-Infrared Flux Ratio}\label{mid}\protect

The HRS model for dust emission does {\it not} apply to the majority of Seyfert 
galaxies, mostly because of the excess at 25\,$\mu$m that characterizes the
objects in the upper right corner of the Seyfert panel in 
Fig. \ref{fig:smg}\footnote{Large values of the $f_\nu(25\mu)/f_\nu(60\mu)$
flux ratio were used initially to define potential Seyfert samples
(de Grijp, Miley, \& Lub \cite{degrijp}).}.
Indeed, the mid-infrared flux ratio
$f_\nu(25\mu)/f_\nu(60\mu)$ in the 12\,$\mu$m Seyferts correlates 
extremely well with $\Gamma$, but not with $L_{FIR}/L_B$ or with 
$L_{FIR}$, unlike optically-selected Seyferts where smaller 
$f_\nu(25\mu)/f_\nu(60\mu)$ is associated with higher $L_{FIR}$
(Hunt \cite{hunt}).
The large deviation from the dust model in Fig. \ref{fig:smg}
must therefore arise from the mid-infrared excess observed in Seyferts
(Miley, Neugebauer, \& Soifer \cite{miley}; Edelson \& Malkan \cite{em}),
attributed to an AGN, either emission directly from
the nucleus or from dust heated by it.

We have also examined the ``warm'' and ``cold'' fractions of the 12\,$\mu$m 
galaxies on the basis of the mid-infrared flux ratio $f_\nu(25\mu)/f_\nu(60\mu)$.
Following Soifer et al. (\cite{soifer2}), we distinguish between
``warm'' and ``cold'' with $f_\nu(25\mu)/f_\nu(60\mu)$\,=\,0.17, and
find 21\% of the normal 12\,$\mu$m galaxies to be warm,
as opposed to the 16\% fraction in the 60-$\mu$m selected BGS.
This higher percentage is almost certainly due to the 12\,$\mu$m flux
selection criterion because of the correlation between $f_\nu(12\mu)$ and $f_\nu(25\mu)$.
The warm fractions of HII/starbursts (18\%) and LINERs (13\%) are similar 
to the BGS, while the much greater fractions (almost 60\%) fractions 
of warm Seyferts arise from the mid-infrared excess cited above.

We have investigated the occurrence of rings and bars in warm and cold
galaxies in the E12GS. 
It turns out that the ring fractions in HII/starbursts and LINERs are
similar for warm and cold objects, but the outer ring fraction in
warm normal 12\,$\mu$m galaxies is greater (17\%) than in cold ones (9\%),
a significant trend at 3.1-$\sigma$. 
A similar conclusion holds for the outer rings in Seyferts:
54\% (50\%) of warm Seyfert 1s (2s) have outer rings, versus 33\% (18\%) of cold
Seyferts, but only the difference in Seyfert 2s is --marginally-- significant
at 1.9\,$\sigma$.
It therefore seems plausible to interpret the
higher outer ring fractions in warm normal 12\,$\mu$m galaxies 
as due to unidentified Seyferts, seen also in the Normal panel
in Fig. \ref{fig:smg} as the vertically-displaced outliers. 
There are no significant differences in the bar frequencies between 
warm and cold 12\,$\mu$m galaxies of any activity class.
We conclude then that even though 1/5 of the E12GS is 
``warm'', as opposed to 1/6 of the 60-$\mu$m selected BGS, the
ring and bar fractions do not depend significantly on mid-infrared
flux ratio, except perhaps as it relates to Seyfert activity.

\subsection{Star Formation Summary}

The analysis of the infrared properties of the E12GS reveals that:
\begin{enumerate}
\item
The $L_{FIR}/L_B$ ratios of galaxies in the E12GS are 
intermediate between optically-selected and 60\,$\mu$m-selected
samples, but tend to be higher in the HII/starbursts and in the Seyfert 2s.
\item
While the high values of $L_{FIR}/L_B$ and the general
properties of the infrared emission in 12\,$\mu$m normal galaxies,
LINERs, and HII/starbursts can be explained by the two-phase
dust model proposed in HRS, those of Seyferts cannot.
\item
Deviations from the dust model in Seyferts seems to be due to their
excess emission at 25\,$\mu$m, thought to have origin in either
dust heated by the AGN, or in the AGN itself.
\item
Bar fractions in the E12GS are significantly greater for greater far-infrared
flux ratio $f_\nu(60\mu)/f_\nu(100\mu)$
{\it only for the normal galaxies}; no class shows a dependence of
bar fractions on $f_\nu(25\mu)/f_\nu(60\mu)$.
\item
Outer rings in normal galaxies and Seyferts tend to prefer 
higher $f_\nu(60\mu)/f_\nu(100\mu)$ and $f_\nu(25\mu)/f_\nu(60\mu)$\footnote{The
former significantly for Seyfert 1s, while the latter for Seyfert 2s.},
while for inner rings the reverse is true. 
Nevertheless, the high outer ring fractions in Seyferts and the
high inner ring fraction in LINERs do not appear to be the
result of a selection artifact.
\end{enumerate}

These considerations lead us to conclude that the 12\,$\mu$m selection
criterion does influence the star-formation activity of the E12GS, but the effect
is mild compared to that of 60\,$\mu$m selection.
Not unexpectedly, the 12\,$\mu$m HII/starbursts display extreme
values of the ``star-formation'' indicators, but they are compatible
with trends predicted by two-component dust models. 
The infrared properties of the 12\,$\mu$m Seyferts may be partly affected by
recent star formation activity, at least in Seyfert 2s,
but there is undeniably a strong contribution from an AGN, 
since $\Gamma$ correlates very well with the mid-infrared
flux ratio $f_\nu(25\mu)/f_\nu(60\mu)$, known to be an indicator of Seyfert activity.

\section{Interpretation and Implications}\label{interpretation}\protect

We have examined the morphological attributes and global star-formation
characteristics of the 891 galaxies in the E12GS.
In what follows, we attempt to link together some of the results, and
provide a coherent picture of what they might imply for 
galaxies hosting AGNs and starbursts in particular, 
and for spiral galaxies in general.

\subsection{How Many AGN Do We Miss from Extinction in the Galactic Disk?}
If we simply assume that the {\it intrinsic} distribution of $b/a$ in
the Seyfert host galaxies must be uniform, like what we observe in
the normal 12\,$\mu$m galaxies, we can estimate how many inclined Seyferts
are missing from the 12\,$\mu$m sample.
Even if {\it all of the Seyferts} within the flux limit were picked up
by the E12GS, the identification of their Seyfert nuclei still rests on
optical spectroscopy, which could be incomplete in the presence of
large extinction in the body of the host galaxy.
The $b/a$ distributions of the 12\,$\mu$m Seyfert 1s and 2s can be made
identical to that of the normal galaxies with the addition
of 5 highly inclined ($b/a\,<\,$0.4) Seyfert 1s, and 5 highly inclined Seyfert 2s.
This corresponds to incompletenesses of 12\% and 9\% for
the Seyfert 1s and 2s, respectively.
This may be however an overestimate, 
especially for type 1s with a median Hubble type of Sa,
since the axial-ratio distributions of the 12\,$\mu$m Seyferts are 
very similar to those of normal early-type spirals 
(Binney \& de Vaucouleurs \cite{binney:dev}). 

\subsection{Why are Outer Rings but not Bars Abnormally Frequent in Seyfert Nuclei?}
Previous researchers suggested that, since many reasonably-sized
galaxies already harbor massive black holes, perhaps from their
early formation days, the key to making them Seyferts is a gas
supply that can fuel their nucleus during the current epoch.
Alternatively, black holes could be formed and maintained by normal stellar 
evolution in a massive compact nuclear star cluster
(Norman \& Scoville \cite{norman:scoville}).
Either way, any dynamical mechanism that can redistribute angular
momentum, and cause gas to spiral in very close to the center,
should be strongly associated with central bursts of
star formation and observed nuclear activity.
 
Bars, or nested bars, are thought to be such a mechanism.
In fact they probably {\it do} function in this way,
because they seem to be helpful in promoting a burst of
star formation in the center of a ``normal" galaxy,
as indicated in $\S$\,\ref{sample}.
However, it is not understood why bars do {\it not}
promote Seyfert nuclear activity, even in any weak statistical sense.

One answer to this dilemma may lie in the large fraction of
outer rings in Seyferts.
Buta \& Combes (\cite{bc}) discuss the observational and
theoretical properties of rings in spiral galaxies, and we
mention here a few points germane to our discussion but
not already mentioned previously.
First, bars in galaxies require between 2 and 5\,$\times\,10^8$~yrs to form
(Combes \& Elmegreen \cite{combes:elmegreen}); subsequently,
formation of nuclear and inner rings requires roughly $10^8$~yrs, and
outer rings about 3$\,\times\,10^9$~yrs. 
Second, while outer rings ({\it R}) can be sustained for a Hubble time in the
absence of tidal interactions, in dense environments they are relatively 
fragile and tend to be either completely destroyed or converted into
pseudo-rings ({\it R$^\prime$}).
Third, inner or nuclear rings can have a lifetime as short as $10^8$~yrs, 
because of nuclear gas inflow and consequent star formation.  
Fourth, because of the long time necessary for the
formation of an outer ring, true outer rings\footnote{As opposed to
pseudo-outer rings which appear to be younger.}
would not
be expected to be observed preferentially in strongly barred galaxies.
This last is true because strong bars are thought to be only a
transient phase in the life of the galaxy, as they can be
dissolved or converted into lenses or triaxial configurations
by massive gas flow to the center. 
It follows that rings may supplant bars as signals of historical angular
momentum transfer late in the life of the galaxy.
Seyfert galaxies, with their high ring frequency, may have reached
an advanced stage in their evolution which would be characterized 
by ``older'' indicators of angular-momentum transfer. 

Such timescale considerations would also help explain the
extremely high bar fraction in HII/starburst galaxies.
If bars, associated with centralized starbursts, signal mass
transfer {\it only early on in the galaxies' development}, then
we could interpret the low outer-ring fraction in HII galaxies
as an indication of lack of time; outer rings have not yet had
time to form in starbursts.
On the other hand, 80\% (of the 20\%) {\it unbarred} HII fraction present
inner rings; such structures would have had sufficient formation time 
if the timescales delineating ``early'' and ``late'' (or young
and old) are roughly $10^8$ and $10^9$~yrs, respectively.
An implication of this scenario is that 
the ``trigger delay time" for Seyfert activity should be much
longer than that for starbursts. 

\subsection{Implications for AGN Unification}
The core idea of AGN Unification schemes is that apparently different
types of nuclear activity may in fact be intrinsically similar, if only
we could observe them more fully.
Two examples of AGN unification are the hypotheses that:
{\it i}) the emission lines in ``classical" LINERs are photoionized by
a hard continuum, the same as in Seyfert 2 nuclei, except that
the ionization parameter is lower because of the lower intensity
of radiation reaching the emitting clouds; or
{\it ii}) each narrow-line AGN (e.g. Seyfert 2) actually contains a broad-line
region like those in Seyfert 1s, but something obscures it from our view.
 
One 
implication of the unification hypotheses is that
the apparently different types of active nuclei should reside in
statistically similar galactic environments. In other words, the host
galaxy should not ``make much difference" to the AGN, if they are all
fundamentally the same kind of objects.  Thus a (negative) test
is proposed: are there any significant differences between the host
galaxies of different classes of AGNs?
As usual, the most difficult aspect of such tests is finding
suitably unbiased samples, so that apparent differences are not 
artificially produced by selection effects.
 
This was a strong motivation for our morphological study of the
host galaxies of the 12\,$\mu$m AGN. In terms of bars, the 12\,$\mu$m AGN pass
the unification test, since we could not find any significant differences
among AGN types.  In contrast, the HII-region galaxies show a higher bar
frequency.  In this regard, then, LINER galaxies resemble more the
Seyferts than the HII/starbursts. 
 
In our relatively unbiased AGN sample,
we found apparent host-galaxy differences among different AGN types:
{\it i}) The LINERs appear to have an unusually high incidence of
{\it inner} ($\sim\,10^8$~yrs formation time) rings, in contrast to the Seyfert 1s which 
have an unusually high incidence of {\it outer} ($\sim\,10^9$~yrs formation time) rings.  
This complicates the simplest version
of unification for LINERs and Seyferts. One possible embellishing solution,
for example, might be to postulate that LINER and Seyfert nuclei
are the same objects, but seen at different evolutionary stages,
as revealed by the differences in the rings in their host galaxies
(see above).
{\it ii}) The Seyfert 2s appear to have later morphological types than 
Seyfert 1s. Again a possible complication 
could be that Seyfert 2 nuclei are dustier, creating more obscuration
along more lines-of-sight (see discussion in MGT).

We could carry the proposed negative test further by formulating a 
counter-hypothesis: Seyfert, LINER, and starburst activity in galaxies
could be directly related to the evolutionary status of the galaxy, 
at least since its last major disturbance.
Previous work (see Introduction) suggests that non-axisymmetric
perturbations, such as bars, induce rapid evolution of spiral disks
through transfer of angular momentum;
evolution in this sense is most likely from late to early 
spiral types (Pfenniger \cite{pfenniger}; Martinet \cite{martinet}).
The presence of gas is fundamental in this process because of
its dissipative properties and destabilizing influence.

It follows that Seyfert nuclei, hosted by predominantly early-type spirals, 
would represent a more evolved manifestation of activity (central gas deposit) 
than either LINERs or starbursts which tend to be found in later Hubble 
types.
Type 1 Seyferts would also be older, more evolved, than type 2s
since they are found in slightly earlier morphological types. 
The high frequency of outer rings in Seyferts, inner rings in LINERs,
and bars in HII/starbursts
would also follow, since the different formation times and lifetimes
of these structures would reflect differences in the evolutionary stage of 
the galaxies.
If peculiar morphologies reflect recent (young) disturbances, likely since they
occurred more frequently in the past (van den Bergh \cite{vdb}, and
references therein), then the peculiar fractions that decrease systematically 
going from HII/starbursts (45\%), to LINERs (37\%), Seyfert 2s (27\%), and 
Seyfert 1s (21\%) obey a similar trend in evolutionary status.
The high HI mass fractions in starbursts that decrease going to Seyfert
2s and 1s, together with an opposing trend in disk surface brightness
(Hunt et al. \cite{hunt2}) are also compatible with such a
progression of activity and Hubble type.
Finally, an evolutionary sequence from starbursts to Seyfert 2s, to type 1s 
(e.g., Oliva et al. \cite{oliva}) would be a possible consequence.

\begin{figure}
\centerline{\epsfig{figure=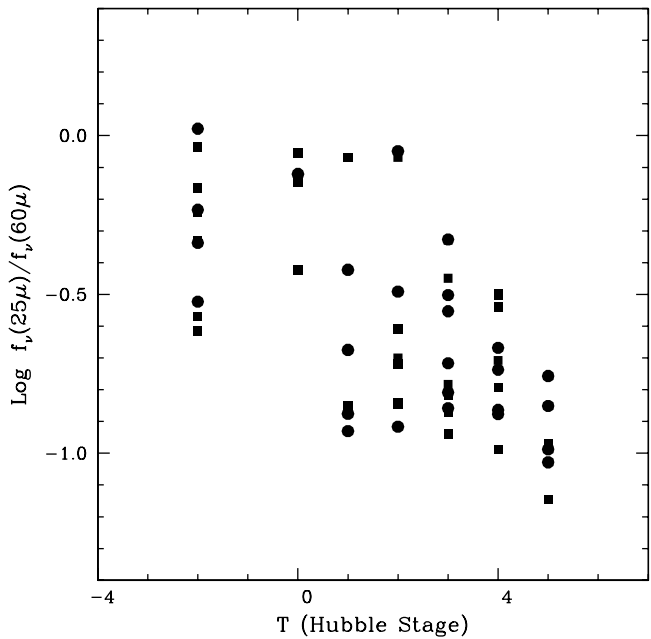,width=10cm}}
\figcaption[morphregres.ps]{Plots of mid-infrared flux ratio 
$f_\nu(25\mu)/f_\nu(60\mu)$ versus Hubble type index T for the 
12\,$\mu$m Seyferts.
Seyfert 1s are marked with filled circles, and Type 2s with filled squares.
The parametric correlation coefficient for the regression is -0.62
for 28 data points, corresponding to a Student-$t$ statistic of 4.06.
\label{fig:agn}
}
\end{figure}

In the same vein, if Seyfert 2 nuclei were younger/less evolved than type 1s,
and considering that type 2 nuclei tend
to be weaker relative to the galaxy than type 1s (Yee \cite{yee}),
we would deduce that the intensity of nuclear activity increases with age.
If age goes hand-in-hand with morphological type, 
we might therefore expect to find trends with measures of nuclear activity.
Figure \ref{fig:agn}, where the mid-infrared flux
ratio $f_\nu(25\mu)/f_\nu(60\mu)$ is plotted against the Hubble type index,
may show such a trend.
As discussed in $\S$\,\ref{mid}, $f_\nu(25\mu)/f_\nu(60\mu)$ is a good
indicator of Seyfert activity, and the figure shows a correlation
between it and Hubble type T; the correlation is significant at $>$\,99.9\%
(two-tailed).
As the mid-infrared flux ratio becomes larger, the morphological types
become earlier, and the nuclei older and more intense, if
time increases to the left as it would were our hypothesis correct.
If $f_\nu(25\mu)/f_\nu(60\mu)$ measures black hole mass,
presumably related to nuclear intensity,
such a trend would also be consistent with 
the correlation between black hole mass and total bulge luminosity
(Kormendy \& Richstone \cite{kormendy:richstone}),
since early-type spirals tend to have more luminous bulges than late types.

While our findings are inconsistent with some of the consequences of the
Unification Scheme for AGNs, and perhaps more consistent with other
hypotheses, 
our study suffers from limitations.
We have used qualitative data from the literature, and some of the results
are of necessity plagued by small-number statistics.
High-quality multiwaveband image data are needed to systematically quantify the 
occurrence of bars, rings, and lenses,
and to better evaluate their influence on the galaxy as a whole.
Indeed, lenses may follow from bar dissolution (Combes \cite{combes}),
but the lack of consistent literature data in this regard precluded analysis.
A subsequent paper will report on an optical and near-infrared image atlas
of the 12\,$\mu$m Seyferts with the aim of further investigating the
connection between nuclear activity and galaxy morphology. 

\acknowledgements
This research was partially funded by ASI Grant ARS-98-116/22.
One of us (M.A.M.) would like to thank the Italian National Astronomy
Group (G.N.A. -- C.N.R.) and the Osservatorio Astrofisico di Arcetri
for financial support.
Extensive use was made of the NASA/IPAC Extragalactic Database (NED),
operated by the Jet Propulsion Lab, Caltech, under contract with NASA.
We thank Brian Rush for providing tabulations of some of the IRAS data,
and Roberto Maiolino for stimulating discussions.
Finally, we are grateful to the referee Susan Simkin for helpful
suggestions and insightful comments.

\newpage

\end{document}